\documentclass[%
 amsmath,amssymb,
 reprint,%
 floatfix,
]{revtex4-1}

\usepackage{graphicx}
\usepackage{dcolumn}
\usepackage{bm}
\usepackage{float}
\usepackage{amsmath}

\usepackage[utf8]{inputenc}
\usepackage[T1]{fontenc}
\usepackage{mathptmx}
\usepackage{etoolbox}

\makeatletter
\def\@email#1#2{%
 \endgroup
 \patchcmd{\titleblock@produce}
  {\frontmatter@RRAPformat}
  {\frontmatter@RRAPformat{\produce@RRAP{*#1\href{mailto:#2}{#2}}}\frontmatter@RRAPformat}
  {}{}
}%
\makeatother
\begin{document}

\preprint{AIP/123-QED}

\title{Turbulence-resistant self-focusing vortex beams}

\author{Meilan Luo}
\affiliation{Department of Physics and Synergetic Innovation Center for Quantum Effects and Applications, Hunan Normal University, Changsha, Hunan 410081, China}
\affiliation{Institute of Photonics, University of Eastern Finland, P.O. Box 111, FI-80101 Joensuu, Finland}
\author{Matias Koivurova}
\affiliation{Tampere Institute for Advanced Study, Tampere University, 33100 Tampere, Finland}
\affiliation{Faculty of Engineering and Natural Sciences, Tampere University, 33720 Tampere, Finland}
\email{matias.koivurova@tuni.fi}
\author{Marco Ornigotti}
\affiliation{Faculty of Engineering and Natural Sciences, Tampere University, 33720 Tampere, Finland}
\author{Chaoliang Ding}
\affiliation{Department of Physics and Henan Key Laboratory of Electromagnetic Transformation and Detection, Luoyang Normal University, Luoyang 471934, China}

\date{\today}

\begin{abstract}
We consider recently introduced self-focusing fields that carry orbital angular momentum (OAM) [Opt. Lett. \textbf{46}, 2384--2387 (2021)] and in particular, their propagation properties through a turbulent ocean. We show that this type of field is especially robust against turbulence induced degradation, when compared to a completely coherent beam. In moderately strong oceanic turbulence, the self-focusing OAM beam features over five orders of magnitude higher peak intensities at the receiver plane, an $\sim$80 \% detection probability for the signal mode, as well as an energy transmission efficiency in excess of 70 \% over a link of $\sim$100 m. Counter-intuitively, the focusing properties of such fields may be enhanced with increasing turbulence, causing the mean squared waist to become smaller with greater turbulence strength. Our results demonstrate that certain types of partial coherence may be highly desirable for optical telecommunication employing OAM.
\end{abstract}

\maketitle

\section{Introduction}
Structured light, and in particular, light carrying orbital angular momentum (OAM) has been the subject of intense research since the seminal work by Allen and Woerdman in 1992 \cite{allen}. Several fields of fundamental and applied research, such as microscopy \cite{hell}, spectroscopy \cite{noyan}, particle manipulation \cite{singh}, quantum information \cite{erhard}, and optical communications \cite{commun}, may benefit from light that carries OAM. A comprehensive review of the field, and a discussion of its future challenges can be found in Refs. \citenum{andrews} and \citenum{roadmap}, respectively. Amongst the different possibilities offered by structured light, vortex beams are the most promising candidates for the realisation of optical communication protocols, since they possess a virtually infinite alphabet that allows dense coding, multiplexing, and high-bit-rate communications \cite{wang1,gong1}. Vortex beams possess a twisted wave front, at the core of which sits a topologically-protected phase singularity described by the characteristic helical wavefront $e^{im\phi}$, where $\phi$ is the azimuthal angle around the beam propagation axis. The index $m\in\mathbb{Z}$ is known as the topological charge, which defines the amount of OAM carried by the beam, i.e., $\hbar m$ \cite{andrews}. The unbounded nature of the index $m$ translates to an infinite dimensional Hilbert space available to OAM states -- as opposed to the bounded two-dimensional Hilbert space associated to polarisation -- which is ultimately what makes OAM enticing for optical communication purposes.

One of the main limiting factors that prevents OAM-carrying beams to be efficiently employed for free space optical communication, is the signal degradation due to atmospheric turbulence \cite{turbulence1,turbulence2}. For instance, one significant challenge that needs to be resolved is the modal cross-talk caused by turbulence, which can dramatically impair system performance \cite{paterson, tyler}. The dispersion of the OAM spectrum comes from the aberrant wavefront distorted by turbulence. The stronger the turbulence is, the more the spectrum will disperse \cite{fu}. Additionally, vortex beams experience significant divergence upon propagation -- which is made worse by turbulence -- making detection difficult due to limitations in detector size at the receiver plane \cite{metrics}. Mitigating the effects of turbulence is an active area of study, and several schemes have been proposed to overcome these limitations, including adaptive-optics-enabled turbulence compensation \cite{ren}, Laguerre-Gaussian mode sorters \cite{Fontaine}, and even deep learning \cite{deeplearning}.

However, all the cases mentioned above consider the propagation of a fully coherent beam through the atmosphere. It is widely known that partially coherent fields feature superior resilience to intensity signal degradation upon propagation through turbulent media \cite{wuj, gburmedia, ricklin, aristide,scintillation1}. These results hint at the possibility of utilising partially coherent vortex beams to implement free space optical communication channels. Indeed, it has been revealed that one can reduce the turbulence-induced scintillation by decreasing the coherence of a vortex beam \cite{scintillation2}, which comes at a price of increased power loss over the link. Meanwhile, other studies have suggested that partially coherent vortex beams perform \textit{worse} in turbulence than their coherent counterparts \cite{liu, channel}. These studies considered Schell-model type correlations (i.e. correlations that depend only on the distance between points), which is very similar to the effect of turbulence to begin with. Partially coherent electromagnetic fields may also feature locally varying degree of coherence, such as in the case of nonuniformly correlated fields \cite{lajunen}, which feature self-focusing upon propagation. It has been shown that self-focusing beams possess better resilience against turbulence-induced noise than fully coherent fields \cite{scintillation1,ding}.

In the present work, we consider a recently introduced class of partially coherent self-focusing fields carrying OAM \cite{mei} and show how the combination of the innate resilience of self-focusing and information carrying capacity of OAM makes them a viable candidate for the realization of an efficient, free space optical communication link. In particular, we show that this kind of vortex beam exhibits self-focusing both in free space and in oceanic turbulence. In fact, it is found that the self-focusing property may be more pronounced in stronger turbulence and that the normalized OAM density shows also focusing behavior, compared to the coherent counterpart. This translates to a beam with several orders of magnitude higher amount of energy at the receiver, combined with an increased detection probability of the sent OAM state.

\section{Self-focusing vortex beams}

To start our analysis, let us consider a statistically stationary scalar source located in the plane $z=0$, radiating a beam-like field towards the positive half-space $z>0$. The spatial coherence properties of a 2D source at points $\boldsymbol{\rho}'_1 = \left(x_1^\prime,y_1^\prime\right)$ and $\boldsymbol{\rho}'_2 = \left(x_2^\prime,y_2^\prime\right)$, at angular frequency $\omega$, can be described by the cross-spectral density (CSD) function \cite{Mandel1995}
\begin{align}
    W_0(\boldsymbol{\rho}'_1,\boldsymbol{\rho}'_2,\omega) = \left\langle E_0^*(\boldsymbol{\rho}'_1,\omega ) E_0(\boldsymbol{\rho}'_2,\omega ) \right\rangle,
\label{D-1}
\end{align}
where $E_0(\boldsymbol{\rho}',\omega)$ is the complex electric field, asterisk denotes the complex conjugate, 
and the angular brackets stand for ensemble averaging. In what follows, the frequency dependence of all quantities of interest will be left implicit for brevity of notation.

It is well-known from standard coherence theory, that for a CSD to be physically realizable, it must be expressible in the form~\cite{Gori}
\begin{align}
    W_0(\boldsymbol{\rho}'_1,\boldsymbol{\rho}'_2) = \int_{-\infty}^\infty p(v) H_0^\ast(\boldsymbol{\rho}'_1,v) H_0(\boldsymbol{\rho}'_2,v){\rm d}v,
\label{D-2}
\end{align}
where $p(v)$ is a non-negative, $\text{L}^2(\mathbb{R}^2)$-integrable probability density function, and $H_0(\boldsymbol{\rho}', v)$ is a positive semidefinite kernel which we can choose as
\begin{align}
    H_0(\boldsymbol{\rho}', v)=\tau (\boldsymbol{\rho}')\exp\left(-2\pi iv{{\rho}'^{2}}\right),
\label{D-3}
\end{align}
without loss of generality. Here $\tau(\boldsymbol{\rho}')$ is a (possibly complex valued) amplitude profile function and $\rho' = \left|\boldsymbol{\rho'}\right|$. The corresponding CSD can then be expressed as
\begin{align}
    W_0(\boldsymbol{\rho};_1,\boldsymbol{\rho}'_2) = \tau^*(\boldsymbol{\rho}'_1) \tau(\boldsymbol{\rho}'_2) \mu_0(\rho'^2_1-\rho'^2_2),
\label{D-4}
\end{align}
where $\mu_0$ is the (generalized) complex degree of spectral coherence, corresponding to the Fourier transform of $p(v)$. Note that the profile function $\tau(\boldsymbol\rho_{1,2}')$ is responsible only for the resulting intensity distribution, whereas the weight function $p(v)$ produces the correlations. For the special case of $v=0$, the above expression reduces to the CSD of a fully coherent field, which we will take as a reference, to benchmark the performance of the self-focusing OAM beams in the next sections. It is also worth mentioning, that since the only assumption on the weight function $p(v)$ is to be square integrable, i.e., Fourier-transformable, regardless of the explicit form of $p(v)$, Eq.~(\ref{D-4}) will always produce a self-focusing field.

From now on we make the explicit choice for the function $p(v)$ to be a Gaussian,
\begin{align}
    p(v) = \sigma^2\sqrt{\pi}\exp\left( -\pi^2\sigma^4 v^2 \right)
\label{D-5}
\end{align}
where $\sigma$ indicates the coherence width of the beam. Beams with a larger value of $\sigma$ have longer focal distances, which may be even on the order of kilometers \cite{ding}. Furthermore, we take the intensity profile of a Laguerre Gaussian beam with radial index zero \cite{andrews,siegman}
\begin{align}
   \tau (\boldsymbol{\rho }') = \left(\frac{\rho'}{w_0}\right)^{|m|} \exp(im\phi) \exp\left( -\frac{\rho'^2}{2w_0^2} \right),
\label{D-8}
\end{align}
where is the transverse beam width at $z=0$, and $m$ is the topological charge. By inserting this into Eq.~(\ref{D-4}), we end up with
\begin{align}
    W_0(\boldsymbol{\rho}'_1, \boldsymbol{\rho}'_2) & = \left(\frac{\rho'_1\rho'_2}{w_0^2}\right)^{|m|} \exp[im (\phi_2 - \phi_1)] \nonumber \\
    & \exp \left( -\frac{\rho'^2_1 + \rho'^2_2}{2w_0^2} \right) \exp\left[-\frac{(\rho'^2_2 - \rho'^2_1)^2}{\sigma^4} \right],
\label{D-9}
\end{align}
The negative sign in the last exponent, refers to the fact that we now have a converging field. Therefore, Eq. (\ref{D-9}) represents a self-focusing vortex beam, carrying $m$ units of OAM.

\section{Propagation in oceanic turbulence}

Under the paraxial approximation, the propagation of a partially coherent beam from the source plane $z=0$ to an arbitrary plane $z>0$ in a turbulent medium can be described by the extended Huygens--Fresnel integral \cite{Mandel1995, Andrews2}
\begin{align}
    W(\bm{\rho}_1, \bm{\rho}_2,z) & = \left( \frac{k}{2\pi z} \right)^2 \iint_{-\infty}^\infty W_0(\boldsymbol{\rho}'_1,\boldsymbol{\rho}'_2) \nonumber \\
    & \hspace{-5mm} \times \exp \left[ -ik\frac{\left(\bm{\rho}_1 - \boldsymbol{\rho}'_1 \right)^2-\left(\bm{\rho}_2 - \boldsymbol{\rho}'_2 \right)^2}{2z} \right] \nonumber\\
    & \hspace{-5mm} \times \left\langle \exp \left[ \varphi (\bm{\rho}_1,\boldsymbol{\rho}'_1,z) + \varphi^\ast (\bm{\rho}_2,\boldsymbol{\rho}'_2,z) \right] \right\rangle_{\rm M} {\rm d}^2 \boldsymbol{\rho}'_1 {\rm d}^2 \boldsymbol{\rho}'_2,
\label{D-10}
\end{align}
where $\bm{\rho}_1 = \left(x_1,y_1\right)$ and $\bm{\rho}_2 = \left(x_2,y_2\right)$ represent two arbitrary spatial positions in the target plane, $k=2\pi/\lambda$ is the wave number, $\varphi(\bm{\rho},\boldsymbol{\rho}', z)$ denotes the phase perturbation induced by the refractive-index fluctuations of the random medium between $\bm{\rho}'$ and $\boldsymbol{\rho}$, and $\left\langle...\right\rangle_{\rm M}$ is the ensemble average over M realizations of turbulent media. Here, we consider oceanic turbulence to find the performance of the beam in the worst case conditions.

If the fluctuations of the medium are homogeneous and isotropic, then the ensemble average in Eq.~(\ref{D-10}) can be expressed analytically as
\begin{align}
    & \langle \exp \left[\varphi^\ast(\bm{\rho}_1,\boldsymbol{\rho}'_1,z)+\varphi(\bm{\rho}_2,\boldsymbol{\rho}'_2,z)\right]\rangle_{\rm M} =  \exp\left\{ -\frac{1}{3}\pi^2 k^2zT \right. \nonumber \\ 
    & \hspace{10mm} \left. \left[ (\bm{\rho}_1 - \bm{\rho}_2)^2  + (\bm{\rho}_1-\bm{\rho}_2) \cdot (\boldsymbol{\rho }'_1 - \boldsymbol{\rho }'_2)+(\boldsymbol{\rho }'_1 - \boldsymbol{\rho }'_2)^2 \right] \right\},
\label{D-11}
\end{align}
where $T$ is the turbulence parameter, which, for the case of oceanic turbulence, has the following explicit form \cite{Thorpe}
\begin{align}
    T & =  0.388\times 10^{-8} \varepsilon^{-1/3} \chi_T  \nonumber \\ 
    & \times \left( 47.5708\varpi^{-2} - 17.6701\varpi^{-1} + 6.78335\right).
\label{D-14}
\end{align}
In the above equation, $\varepsilon$ is the rate of dissipation of turbulent kinetic energy per unit mass of fluid, which may vary in the range [$10^{-1},10^{-10}~{\rm m}^{2}/{\rm s}^{3}$], $\chi_T$ is the rate of dissipation of mean-square temperature, which has values in the range $[10^{-4}, 10^{-10}]~{\rm K}^2/s$. Furthermore, the (dimensionless) parameter $\varpi$ denotes the relative strength of temperature and salinity fluctuations, which for ocean water, ranges from $-5$ to $0$, whose limits correspond to dominating temperature-induced and salinity-induced optical turbulence, respectively. The parameter $T$ usually attains values ranging from about $10^{-16}$ for weak turbulence, up to $10^{-12}$, for very strong turbulence. Since the individual parameters in Eq.~(\ref{D-14}) are not relevant to our study, we use the value of $T$ directly, mainly in the moderate to strong region of $10^{-14}$ to $10^{-12}$.

To obtain the expression of the propagated CSD in presence of turbulence, we then make use of Eqs.~(\ref{D-2}) and (\ref{D-10}), and substitute them into Eq.~(\ref{D-11}), such that at an arbitrary plane $z$ we have the following CSD
\begin{align}
    W(\bm{\rho}_1, \bm{\rho}_2,z)=\int_{-\infty}^\infty p(v)H^\ast(\bm{\rho}_1,v,z)H(\bm{\rho}_2,v,z){\rm d}v,
\label{D-15}
\end{align}
where the propagated kernels are given by
\begin{align}
    & H^\ast(\bm{\rho}_1,v,z)H(\bm{\rho}_2,v,z) = 
    \left(\frac{k}{2\pi z}\right)^2 \iint_{-\infty}^\infty H_0^\ast(\boldsymbol{\rho}'_1,v) H_0(\boldsymbol{\rho}'_2,v) \nonumber \\ 
    & \hspace{15mm} \times \exp\left[-ik\frac{\left(\bm{\rho}_1-\boldsymbol{\rho }'_1\right)^2-\left(\bm{\rho}_2- \boldsymbol{\rho}'_2\right)^2}{2z} \right]\nonumber \\
    & \hspace{15mm} \times \langle \exp \left[\varphi^\ast(\bm{\rho}_1,\boldsymbol{\rho}'_1,z)+\varphi(\bm{\rho}_2,\boldsymbol{\rho}'_2,z)\right]\rangle_{\rm M} {\rm d}^2\boldsymbol{\rho}'_1 {\rm d}^2\boldsymbol{\rho}'_2.
\label{D-16}
\end{align}

Upon substituting Eqs.~(\ref{D-3}) and (\ref{D-8}) into Eq.~(\ref{D-16}) and going through some lengthy but straightforward algebra, we obtain the following expression for the product of two propagation kernels of a self-focusing vortex beam
\begin{align}\label{hh}
    & H^\ast(\bm{\rho}_1,v,z)H(\bm{\rho}_2,v,z) = \nonumber \\
    & \hspace{4mm} \exp{\left[-\Theta(x,y,T)-\frac{ik}{2z}\left(\boldsymbol\rho_1^2-\boldsymbol\rho_2^2\right)-T_1\left(\boldsymbol\rho_1-\boldsymbol\rho_2\right)^2\right]}\nonumber\\
    & \hspace{4mm} \times w_m^2(z)\sum_{n=0}^{|m|}\left(\tilde{C}_{|m|}^n\right)^2\left[F(x,y,T)\right]^{|m|-n}.
\end{align}
Here we have employed multiple shorthand notations, all of which can be found in Appendix A together with a detailed derivation of the equation. By employing the propagated kernels for each $v$ together with Eq.~(\ref{D-15}), we can find the correlation properties of the self-focusing vortex beam at any plane of propagation. 

\section{Propagation properties}
\subsection{Spectral density}
We now use the  formulae derived in the previous section to investigate the spectral density properties of the self-focusing vortex beams. When the two points coincide, $\bm{\rho}_1=\bm{\rho}_2=\bm{\rho}$, Eq. (\ref{hh}) describes the spectral density of a single mode, and can be written in a simplified form as
\begin{align}
\label{hh2}
    & S(\bm{\rho},v,z) = \nonumber \\
    & \hspace{2mm} w_m^2(z)\exp\left[-\frac{\bm{\rho}^2}{w^2(z)}\right]
    \sum_{n=0}^{|m|} \left(\tilde{C}_{|m|}^n\right)^2 \left[\frac{kw_{fr}(z)\bm{\rho}}{zw(z)}\right]^{2(|m| - n)}.
\end{align}
For the case $T=0$, i.e., in absence of turbulence, the waist of a single $v$-mode, $w(z)$, reduces to its free space counterpart $w_{fr}(z)$ (see Appendix A). Further, if the OAM index is set to $m=0$, then the intensity distribution reduces to a self-focusing field with no angular momentum, such as in Ref.~\citenum{ding}, i.e., $S(\boldsymbol\rho,v,z)=[w_0/w(z)]^2\exp{-\boldsymbol\rho^2/w(z)^2]}$. The overall spectral density~\cite{Mandel1995} can be retrieved with
\begin{align}
    S(\bm{\rho},z) = W(\bm{\rho},\bm{\rho},z) = \int_{-\infty}^\infty p(v)S(\bm{\rho},v,z){\rm d}v.
\label{D-18}
\end{align}

\begin{figure*}
\centering
    \includegraphics[width=2\columnwidth]{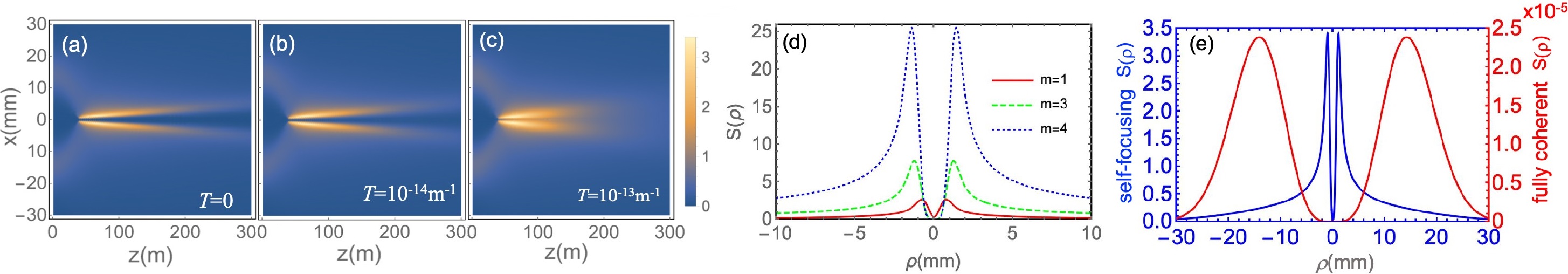} 
     \caption{Spectral density distribution of the self-focusing vortex beam propagating in free space (a) and turbulent ocean (b, c) in the $x-z$ plane with $m=2$. (d) The transverse intensity distributions for various azimuthal index $m$. (e) The transverse intensity distributions of the self-focusing vortex beam and the fully coherent vortex beams with $m=2$. The values for the turbulence parameter used for panels (a)-(c) are indicated on the density plots. For panels (d) and (e), the values $z=60$ m, and $T=10^{-14}$ $\text{m}^{-1}$ have been used.}\label{figure1}
\end{figure*}

In the following, we analytically simulate a beam propagating in oceanic turbulence. The employed parameters are given in the figure captions, whereas the wavelength, $\lambda=632{\rm nm}$, mode waist, $w_0=1{\rm cm}$, and coherence width, $\sigma=5{\rm mm}$ are kept constant. These choices are not the only possible, and they produce a beam with about 60 m focal distance. As the sign of $m$ just changes the handedness of the phase, we consider the case of positive topological charge only. Free space propagation can be regarded as the simplest case where the turbulence term as well as the summation index are both null, i.e. $T = 0$, $n=0$ (see Appendix A).

Figure \ref{figure1} shows the evolution of the spectral density of the self-focusing vortex beam in the $x-z$ plane in free space (a), and oceanic turbulence, for both  weak (b), and strong (c) turbulence. Moreover, panel (d) shows the transverse distribution of the overall spectral density $S(\boldsymbol\rho,z)$ for different values of $m$ at a selected distance $z=60{\rm m}$. The peak value of the intensity and the radius of the dark zone increase for larger value of the topological charge, as expected for OAM-carrying beams \cite{andrews}.
A comparison between self-focusing vortex beam and a fully coherent vortex beam is made in panel (e), to help to evaluate the focusing effect. Here, we show the first main result of our work: the peak intensity of the self-focusing vortex beam is more than five orders of magnitude larger than that of a fully coherent vortex beam, although their intensities are of the same magnitude at the initial plane. This is due to the correlation induced self-focusing.

\subsection{Mean squared beam width}
Let us next evaluate how the width of the beam evolves as it propagates in a turbulent ocean, by defining the mean squared beam width as
\begin{equation}
\label{msw}
w^2_{ms}(z)=\frac{\int_0^{\infty}\bm{\rho}^2 S(\bm{\rho},z){\rm d}\bm{\rho}}{\int_0^{\infty}S(\bm{\rho},z){\rm d}\bm{\rho}}.
\end{equation}
Substituting Eq.~(\ref{hh2}) into the above expression, one can easily see that the behavior of $w_{ms}^2(z)$ is influenced by the OAM order $m$, the turbulence parameter $T$, and the propagation distance $z$, as one would expect. From this, one can determine the distance $z_{min}$, at which $w^2_{ms}$ reaches its minimum. However, since $z_{min}$ involves complicated integrals and summations that do not necessarily admit closed-form solutions, we have to turn to numerical evaluation of the mean squared waist. We first consider a single component of the partially coherent beam, i.e., we take $v$ as fixed. Using Eqs.~(\ref{hh2}) and (\ref{msw}) we can evaluate the mean squared beam width for the $v$-th component of the self-focusing vortex as
\begin{equation}\label{eq25}
w^2_{ms}(v,z) = w^2(v,z)Y_m\left[1-\frac{w^2(v,z)}{w_{fr}^2(v,z)}\right],
\end{equation}
where the $v$ dependence has been written explicitly for clarity, and
\begin{align}
    & Y_m(x) = \nonumber \\
    & \frac{\Gamma\left(\frac{3}{2}+|m|\right)\Gamma\left(\frac{1}{2}-|m|\right){_2F}_1\left(-|m|,-|m|,-\frac{1}{2}-|m|,x\right)}{\pi(-1)^{|m|}{_2F}_1\left(-|m|,-|m|,\frac{1}{2}-|m|,x\right)}
\end{align}
is the OAM dependent correction term, with $\Gamma(n)$ being the Gamma function, and $_2F_1(a,b;c;x)$ the Gauss hypergeometric function \cite{nist}.

\begin{figure}[ht]
\centering
    \includegraphics[width=\columnwidth]{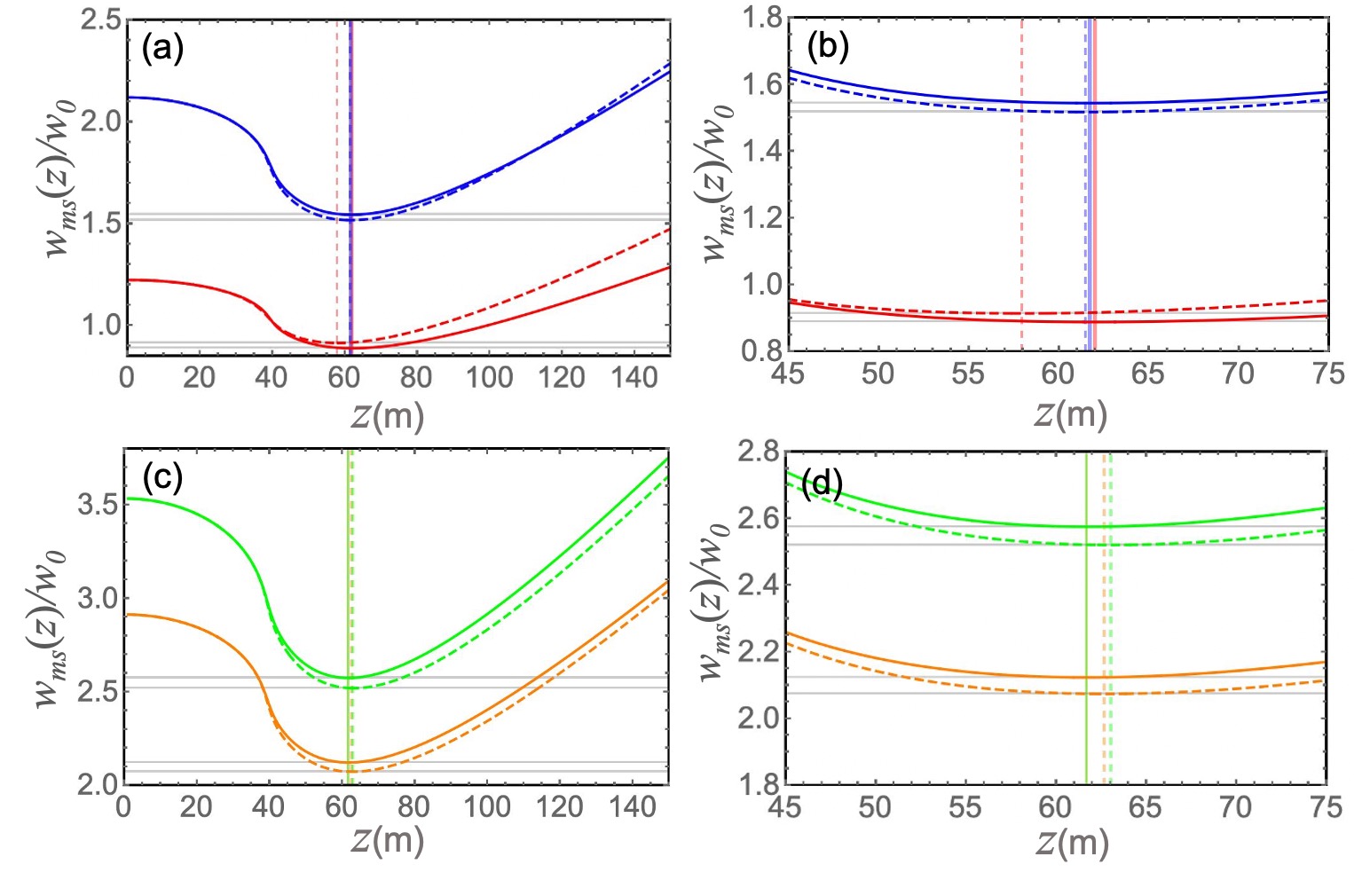} 
     \caption{The root of mean squared width of the self-focusing vortex beam for different values of topological charge and strength of turbulence. Red lines: $m=1$, blue lines: $m=4$, green lines: $m=8$, orange lines: $m=12$. Solid lines in all panels correspond to a value of the turbulence parameter of $10^{-14}$ $\text{m}^{-1}$, while dashed lines correspond to $T=10^{-12}$ $\text{m}^{-1}$. All the other parameters are the same as those used for Fig. \ref{figure1}. }\label{figure2}
\end{figure}

Let us briefly mention some of the main features of the mean squared beam width: (\romannumeral1) For $m=0$, we have $Y_0(x)=1$ for any $x$. Moreover, in free space for any given $|m|>0$ we have $Y_m(0)=1$ as well. (\romannumeral2) Since $w(z)$ increases monotonically with $T$, the mean squared width of the $v$-th component will also monotonically increase with turbulence when $Y_m(x) = 1$. (\romannumeral3) The value of the function $Y_m(-|x|)$ increases with increasing $m$, and thus, high OAM content beams have a larger mean square width. (\romannumeral4) For a fixed $|m|>0$, $Y_m(x)$ decreases monotonically with the increase of turbulence strength, until it tends to a constant value after a sufficient propagation distance. These findings are graphically summarized in Fig.~\ref{figure2}.

It is evident that the mean squared beam width becomes narrower over a certain range of distances, as the turbulence increases (dashed lines in Fig. \ref{figure2}). This becomes more obvious for beams with larger OAM content, such as those depicted in panels (c) and (d). Additionally, one can see how the influence of the term $Y_m(x)$ in Eq.~(\ref{eq25}) affects the focusing dynamics of the beam by modifying the minimum value, as well as the focal distance.

Previous studies of partially coherent self-focusing beams (without vortices) in turbulent ocean has revealed that increasing turbulence causes the focal spot to move towards the source plane \cite{ding}. However, in the case of self-focusing vortex beams, the picture is somewhat different. In this case, the focal spot moves towards to the source plane only for low values of OAM, as it can be seen in Fig. \ref{figure2} (a) and (b) for $m=1$. For higher values of OAM, the position of $z_{min}$ is shifted \textit{away} from the source plane, as shown in Fig. \ref{figure2} (c) and (d) for $m=8$ and $m=12$. 

This constitutes the second main result of our work: by evaluating the mean squared beam width we find that, for higher OAM content and within certain distances, the spread of a partially coherent vortex beam in strong turbulence can be \textit{lower} than in weak turbulence. This is, to the best of our knowledge, the first example of an optical beam possessing such a counter-intuitive property.

\subsection{Orbital angular momentum flux density}

We now proceed to our analysis of the OAM properties of the beam. For a scalar and paraxial partially coherent beam, which has on average no spin angular momentum, the $z$ component of the OAM flux density is of the form \cite{kim}
\begin{align}
\label{L0}
    & L_{\rm orb}(\bm{\rho}, z) = \nonumber \\
    & -\frac{\epsilon_0}{k} {\rm Im} \left[ y_1\frac{\partial}{\partial x_2} W(\bm{\rho}_1,\bm{\rho}_2,z) -x_1\frac{\partial}{\partial y_2} W(\bm{\rho}_1,\bm{\rho}_2,z)\right]_{\bm{\rho}_1=\bm{\rho}_2=\bm{\rho}},
\end{align} 
where $\epsilon_0$ denotes the free space permittivity. The quantity $L_{\rm orb}$ is dependent on the vortex strength, as well as the intensity of the a field.

To have a better understanding of the physical picture, we consider the normalized OAM flux density, which describes the OAM per photon \cite{kim},
\begin{equation}
\label{l}
l_{\rm orb}(\bm{\rho}, z) =\frac{\hbar \omega L_{\rm orb}(\bm{\rho}, z)}{S_p(\bm{\rho}, z)},
\end{equation}
where $S_p(\bm{\rho}, z)=kW(\bm{\rho},\bm{\rho},z)/(\mu_0\omega)$ is the $z$ component of the Poynting vector, with $\mu_0$ being the vacuum permeability. Upon substituting Eqs.~(\ref{D-15}) and (\ref{hh}) into Eq.~(\ref{l}), the normalized OAM flux density can be written as
\begin{equation}
\label{l1}
l_{\rm orb}(\bm{\rho}, z)=\hbar \frac{\int_{-\infty}^{\infty} p(v)S_l(\bm{\rho}, v, z){\rm d}v}{\int_{-\infty}^{\infty} p(v)S(\bm{\rho}, v, z){\rm d}v},
\end{equation}
where
\begin{align}
\label{sl}
    S_l(\bm{\rho}, v, z) & = \frac{w_m^2(z)w^2(z)}{w_{fr}^2(z)}  \exp\left[-\frac{\bm{\rho}^2}{w^2(z)}\right] \nonumber \\
    & \hspace{3mm} \times \sum_{n=0}^{|m|} (\tilde{C}_{|m|}^n)^2 (|m|-n)\left[\frac{kw_{fr}(z)\bm{\rho}}{zw(z)}\right]^{2(|m| - n)},
\end{align}
Using Eqs.~(\ref{l1}) and (\ref{sl}), we can study the behaviour of the normalized OAM flux density as a function of the radial position at any propagation distance of a self-focusing vortex beam carrying $m$ units of OAM.

By comparing the expressions in Eqs.~(\ref{hh2}) and (\ref{sl}), one can readily see that in free space (i.e., $T=0$) Eq.~(\ref{l1}) reduces to the well known constant value of $l_{\rm orb}(\bm{\rho}, z)=|m|\hbar$. Thus, self-focusing vortex beams in free space can be viewed as pure fluid-like rotators for which the OAM content is independent of the radial distance. Therefore, the type of self-focusing vortex beams we consider act like pure coherent vortex modes in free space with respect to their OAM content \cite{swartzlander, gburspie}, since $l_{\rm orb}(\bm{\rho}, z)$ is simply proportional to the topological charge $m$.

When turbulence is present, the OAM flux density attains a nontrivial dependence on radial, as well as axial distance. The effect of turbulence on the OAM flux density of partially coherent self-focusing vortex beams for different $m$ at selected distances are shown in Fig. \ref{figure3} below.

\begin{figure}[ht]
\centering
    \includegraphics[width=1\columnwidth]{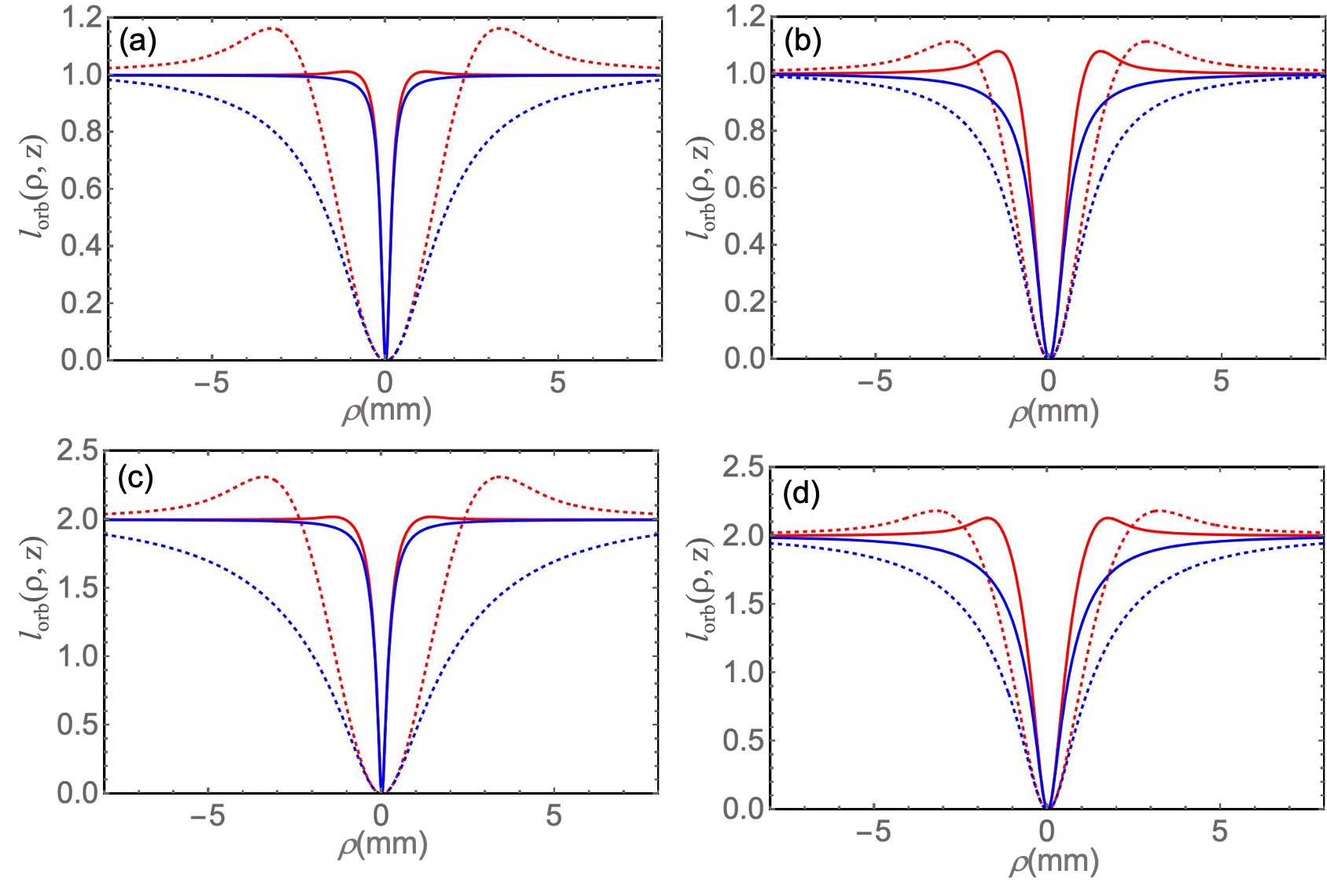} 
     \caption{Distributions of the OAM flux density $l_{\rm orb}(\bm \rho, z)/\hbar$ of self-focusing vortex beams (red lines) and fully coherent vortex beams (blue lines) with $m=1$ (a, b) and $m=2$ (c, d) propagating through oceanic turbulence. Panels (a) and (c) are at a constant propagation distance $z=60{\rm m}$, and solid lines correspond to $T=10^{-14}{\rm m}^{-1}$ while dashed lines correspond to $T=10^{-12}{\rm m}^{-1}$. Panels (b) and (d) are for a constant turbulence parameter $T=10^{-13}{\rm m}^{-1}$, and solid lines correspond to $z=60{\rm m}$ while dashed lines correspond to $z=100{\rm m}$. }\label{figure3}
\end{figure}

Compared to the coherent vortex beams, the self-focusing vortex beams behave more like Rankine vortices, with a small rigid body rotation region near the core (where the OAM depends quadratically on the distance $\rho$) and a fluid-like rotation in the outer regions. With increasing turbulence and propagation distance, the rigid region expands. This is not surprising since, as it follows directly from the last term in Eq.~(\ref{sl}), $\rho$ is inversely proportional to $z$ and $T$. Intriguingly, the maximum value of the OAM flux density for the self-focusing vortex beam exceeds the value of the topological charge carried by the input beam, and it takes a higher value under stronger turbulence. This can be explained as the result of balancing between turbulence induced spreading and the self-focusing of the beam.

\begin{figure*}
\centering
    \includegraphics[width=1.8\columnwidth]{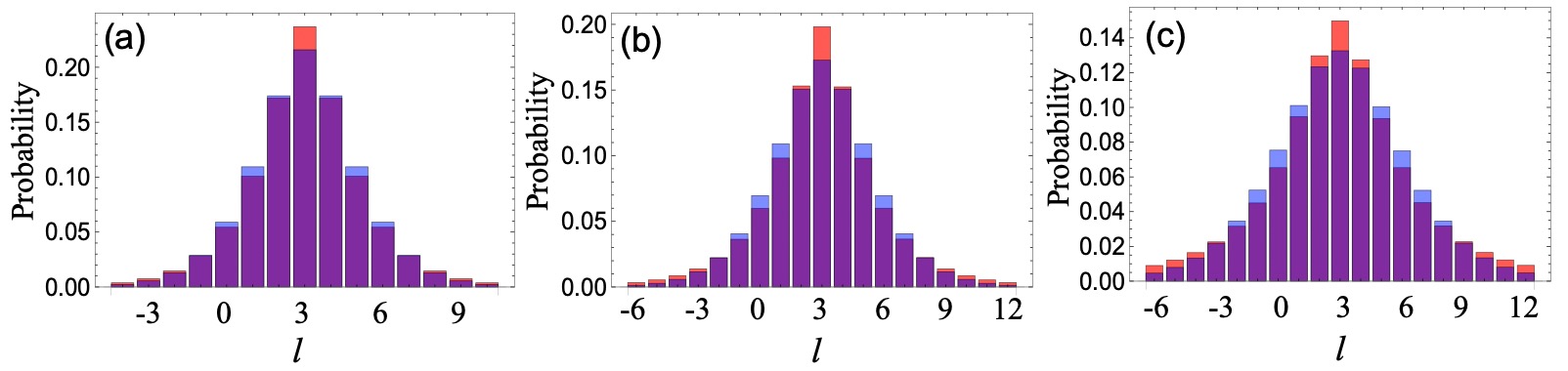} 
     \caption{Probability density for the OAM states of self-focusing vortex beam (red histogram) and fully coherent vortex beam (blue histogram) with $m=3$ through strong turbulence $T=10^{-13}{\rm m}^{-1}$ at different propagation distances with $z=60{\rm m}$ (a), $z=90{\rm m}$(b) and $z=150{\rm m}$ (c).}
     \label{figure4}
\end{figure*}

\subsection{Orbital angular momentum spectrum}
Finally, we investigate the effects of oceanic turbulence on the OAM spectrum of self-focusing vortex beams. It is well known that there is a Fourier relationship between the intensity distribution in the azimuthal direction and the complex OAM spectrum\cite{uncertainty}, i.e., 
\begin{equation}
E(\rho, \theta, z)=\frac{1}{\sqrt{2\pi}}\sum_{l=-\infty}^{+\infty}a_l(\rho, z)\exp(i l\theta),
\end{equation}
and
\begin{equation}
a_l(\rho, z)=\frac{1}{\sqrt{2\pi}}\int_0^{2\pi}E(\rho,\theta, z)\exp(-i l\theta){\rm d}\theta.
\end{equation}
The power weights of partially coherent vortex beams can be written as
\begin{align}
\label{a2}
    & \left\langle|a_l(\rho, z)|^2\right\rangle = \nonumber \\
    & \hspace{10mm} \frac{1}{2\pi}\iint_0^{2\pi}W(\rho,\theta_1,\rho,\theta_2, z){\rm exp}[i l(\theta_1-\theta_2)]{\rm d}\theta_1{\rm d}\theta_2,
\end{align}
in which case the relative power for each OAM mode is defined as
\begin{equation}
\label{P}
P_l(z)=\frac{\int_0^{\infty}\left\langle|a_l(\rho, z)|^2\right\rangle\rho{\rm d}\rho}{\sum_l\int_0^{\infty}\left\langle|a_l(\rho, z)|^2\right\rangle\rho{\rm d}\rho}.
\end{equation}
Computing the relative powers of the states with various azimuthal indices allows us to construct the OAM spectrum \cite{Molina}.

The OAM spectrum of a self-focusing vortex beam, as well as its coherent counterpart for various propagation distances and turbulence parameters are displayed in Fig. \ref{figure4}. Since we saw from the previous subsection that the OAM flux density is a constant in free space for these beams, we can immediately surmise that the OAM spectrum in free space contains only a single peak at $m$ (the signal mode). Therefore, in the following we consider only beams in turbulence.

With increasing propagation distance and turbulence parameter, the energy of the signal mode will inject into the neighbouring modes for both self-focusing and coherent vortex beams, resulting in OAM spectrum dispersion. Thus, the probability of detecting the signal mode decreases monotonously upon propagation through turbulence. The partially coherent self-focusing beam has a heavier tailed OAM spectrum than the fully coherent one. However, the proportion of the signal channel is higher than that of coherent ones, featuring a maximum increase of $\sim$14 \% over the coherent vortex beam. This is partly caused by the narrower beam size of the self-focusing beam, which ensures smaller area affected by turbulence. 

\begin{figure}[ht]
\centering
    \includegraphics[width=1\columnwidth]{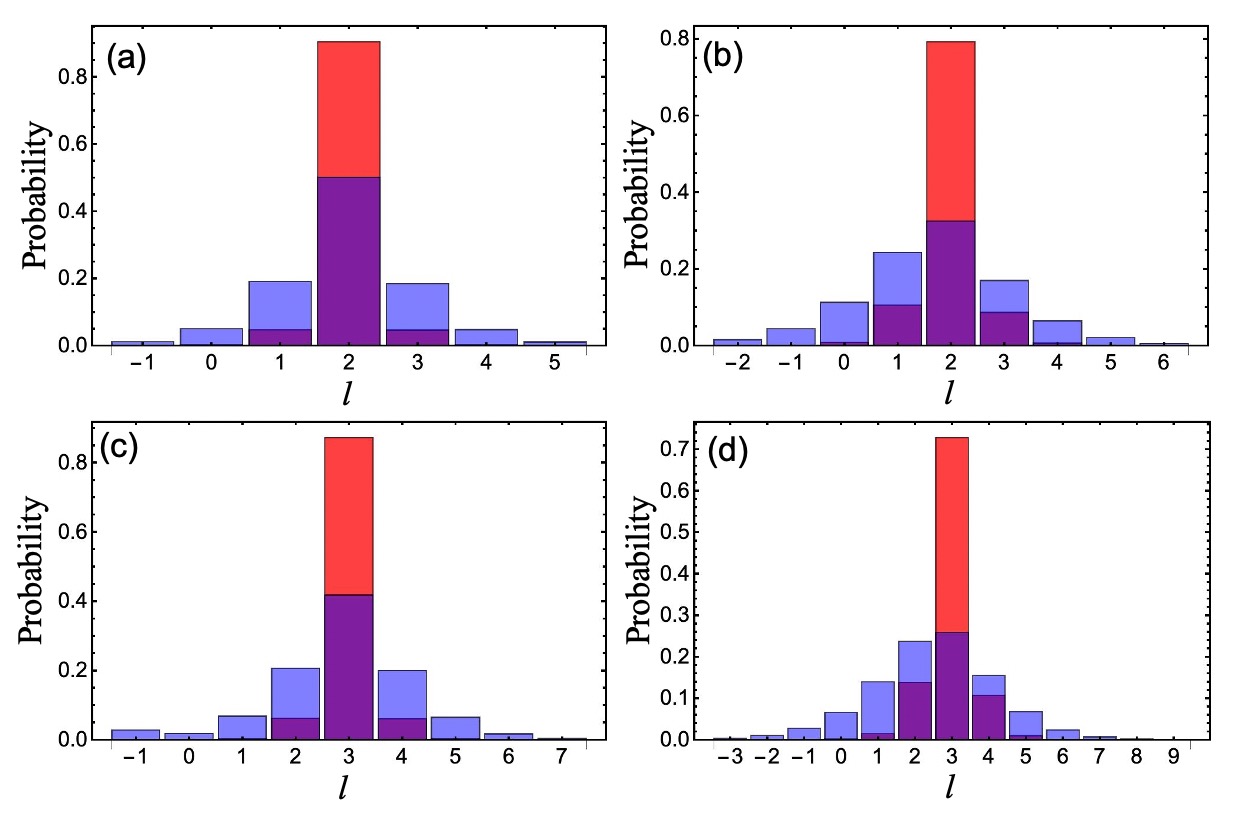}
     \caption{Probability density for the OAM states of self-focusing vortex beam with a finite detector at the receiving plane.  Detector radius is taken to be $10{\rm mm}$, and the initial topological charges are $m=2$ (a,b) and $m=3$ (c,d) at selected propagation distances  with $z=60{\rm m}$ (a, c), and $z=150{\rm m}$ (b, d). Red histogram in all panels corresponds to $T=10^{-14}{\rm m}^{-1}$, while blue histogram corresponds to $T=10^{-13}{\rm m}^{-1}$.}
     \label{figure5}
\end{figure}

As a last example of the present work, we study the OAM spectrum received by a detector at two propagation distances, which is depicted in Fig. \ref{figure5}. Here we assume that the receiver plane may move along the z-axis, and that the employed detector is circular, with a radius of 10 mm. Further, we assume perfect alignment, and thus the detector area contains most of the energy of the self-focusing beam. 

Here, we have the third main result of our work. By comparing Figs \ref{figure4}. and \ref{figure5}., we see that the finite aperture cuts the tails of the OAM spectrum. Hence, the weight of the signal channel becomes much larger, especially for the case of moderately strong oceanic turbulence (>0.8). Furthermore, the proportion of the overall energy that the detector captures from the self-focusing vortex beam with $m=2$ in moderately strong turbulence is $72.3 \%$ at $z=60$ m and $67.7 \%$ at $z=150$ m. In the case of strong turbulence, these values are $72.9 \%$ and $68.9 \% $, respectively. Comparing this to a coherent vortex beam, the ratios at the same propagation distances are $15 \%$ and $14.7 \%$ with $T=10^{-14}{\rm m}^{-1}$, whereas they are $15.1 \%$ and $16.5 \%$ with $T=10^{-13}{\rm m}^{-1}$, respectively. The energy efficiency in the last coherent example is higher in stronger turbulence due to an increase of intensity on axis (i.e. due to vortex breakdown). In each considered case, there is an over four-fold increase in efficiency when compared to a completely coherent beam, and the improvement is greater for longer distances when the focal distance is modified accordingly. 

We can also see that the probability of the signal mode decreases with the increase of topological charge $m$, indicating that a vortex beam with greater $m$ is more vulnerable to turbulence, as larger $m$ gives rise to a wider beam radius, which consequently causes serious OAM dispersion. This means that although it's a good idea to construct high dimensional Hilbert space by increasing the value of topological charge $m$, it may not be efficient in improving free space optical communication. In addition, it is to be noted that the OAM spectrum becomes asymmetric and shifts towards zero under increasing turbulence, which is due to the beam losing angular momentum to the turbulence \cite{Forbes2021}. Mathematically, this can be explained from the turbulence correlated coordinates $a_s$ and $\xi_s$, as the shifts of the coordinates in $x$ and $y$ directions are asymmetric (see Appendix A for more information).

\section{Conclusions}

We have studied the propagation properties of self-focusing vortex beams through oceanic turbulence. We chose ocean environment to find the worst case estimates on the propagation properties of such beams, and the results are applicable also to the much weaker atmospheric turbulence. The self-focusing vortex beams feature several intriguing properties, and they can be extremely robust against turbulence. Although the present work is entirely theoretical, beams such as these may be experimentally generated with deformable optics or digital micro-mirror devices \cite{Yahong}.

First, thanks to the focusing of the beam upon propagation, the peak intensity at the receiver plane is over five orders of magnitude greater than for a corresponding coherent beam. This is an important feature for a beam employed in a data link, since it makes alignment and detection much easier. Here, we considered a beam which has a focal length of about 60 m. However, by increasing the coherence width of the beam, it is possible to get much longer focal distances, even on the order of kilometers. To retain the robustness of the beam, the overall degree of coherence has to remain unchanged, and thus an increase of the coherence width necessitates an increase in the overall beam size. Since the size of the beam is finite in real world applications, this limits the longest achievable focal length. For example, with beam and coherence widths of 20 mm, one can already get a focal length of 0.8 km \cite{ding}.

Second, the width of the beam and the focal distance remain nearly unaltered when the turbulence is increased. In fact, under certain conditions, the mean squared beam waist may become smaller for a larger turbulence parameter. This effect appears to be due to an interplay between OAM and turbulence mediated by the focusing properties of the beam. We believe that we are the first to predict such counter-intuitive behaviour in optical beams.

Third, the signal mode detection probability of the partially coherent beam is improved over the coherent one, and when evaluating OAM spectrum over the whole beam, the signal mode detection probability increases by about 14 \%. When a circular detector of 10 mm radius is employed at the receiving plane, the signal mode is detected with a probability of about 80 \% over moderately strong turbulence. Moreover, the finite detector can collect $\sim$70 \% of the energy that is sent over a distance of $\sim$100 m through strong turbulence, making for an extremely efficient link.

In addition to the properties listed above, partial coherence leads to a significant decrease in scintillation, and therefore improves the receiver’s signal-to-noise ratio. Thus, a data link realized with self-focusing OAM beams has several advantages over coherent OAM beams. The results obtained in the present study may find applications in long distance OAM data links, whether it is through an atmospheric or oceanic environment.

\appendix
\section*{Appendix A}
\renewcommand{\theequation}{A\arabic{equation}}
\setcounter{equation}{0}

In the following, we assume that the initial plane correlation function is of the form of a self-focusing vortex beam that is given in Eq.~(7), the derivation of which was outlined in Sect. 2. of the main text. Further, we assume paraxial propagation towards the positive half-space $z>0$, which is described by the extended Huygens-Fresnel integral of Eq.~(8). By combining these, we attain the kernel propagation integral of Eq.~(12). Upon substituting from Eqs. (3) and (6) into Eq.~(12), we get
\begin{equation}
\begin{split}
    H^*(\bm{\rho}_1, v, z)H(\bm{\rho}_2,v,z) &= 
    \left(\frac{k}{2\pi z}\right)^2\iint \left(\rho' _1\rho' _2\right)^{|m|}e^{-im(\phi_1-\phi_2)} \\
    &\hspace{-12mm}\times {\rm exp}\left(-\frac{\rho'^2_1+\rho'^2_2}{2w_0^2}\right) {\rm exp}\left[-2\pi i v\left(\rho'^2_2-\rho'^2_1\right)\right] \\
    &\hspace{-12mm} \times{\rm exp}\left\{-\frac{ik}{2z}\left[\left(\bm{\rho}_1-{\bm \rho}'_1\right)^2-\left(\bm{\rho}_2-{\bm \rho}'_2\right)^2\right] \right. \\
    &\hspace{-12mm} -T_1\left[\left(\bm{\rho}_1-\bm{\rho}_2\right)^2+(\bm{\rho}_1-\bm{\rho}_2)\cdot ({\bm \rho}'_1-{\bm \rho}'_2) \right.  \\
    &\hspace{-12mm} \left.\left. + ({\bm \rho}'_1-{\bm \rho}'_2)^2\right] \right\} {\rm d}^2{\bm \rho}'_1{\rm d}^2{\bm \rho}'_2,
\end{split}
\end{equation}
where $T_1=\pi^2k^2zT/3$ is the $z$-dependent turbulence parameter. 

To make the integral simpler, we can separate it to ${\bm \rho}'_1$ and ${\bm \rho}'_2$ contributions. The integral over ${\bm \rho}'_1$ takes the following -- analytically solvable -- form in Cartesian coordinates
\begin{equation}
\begin{split}
    & \iint \left[x'_1 - i \text{sgn}(m)y'_1\right]^{|m|} \exp\left[-\alpha_1\left(x'^2_1 + y'^2_1\right) + u_x x'_1 \right.  \\
    &\left. + u_y y'_1 \right]{\rm d}x'_1{\rm d}y'_1  = \frac{\pi}{\alpha_1}\left[\frac{u_x-i\text{sgn}(m)u_y}{2\alpha_1}\right]^{|m|} \exp\left(\frac{u_x^2+u_y^2}{4\alpha_1}\right),
\end{split}
\end{equation}
where $\text{sgn}(x)$ is the sign function. Here we have used the shorthand notations
\begin{subequations}
\begin{gather}
    \alpha_1 = \frac{1}{2w_0^2}-i2\pi v+\frac{ik}{2z}+T_1,
    \\ u_x = \frac{ikx_1}{z}+T_1\Delta x+2T_1x'_2,
    \\ u_y = \frac{iky_1}{z}+T_1\Delta y+2T_1y'_2,
    \\ \Delta x = x_2-x_1, \quad \Delta y=y_2-y_1.
\end{gather}
\end{subequations}
Next, we choose $a_x = kx_1/z-iT_1\Delta x$ and $a_y = ky_1/z-iT_1\Delta y$, so that we can write $u_x=ia_x+2T_1x'_2$, $u_y=ia_y+2T_1y'_2$, then the remaining integral over ${\bm \rho}'_2$ takes the form
\begin{equation}
\begin{split}
    & \iint [ia_x+\text{sgn}(m)a_y+2T_1(x'_2-i\text{sgn}(m)y'_2)]^{|m|} \\ 
    & \hspace{5mm} \times (x'_2+i\text{sgn}(m)y'_2)^{|m|} \exp\left[-\left(\alpha^*_1-\frac{T_1^2}{\alpha_1}\right) \right. \\
    & \hspace{5mm} \times\left(x'^2_2+y'^2_2\right) - \left(\frac{ik}{z}x_2+T_1\Delta x - \frac{ia_xT_1}{\alpha_1}\right)x'_2 \\
    & \hspace{5mm} \left. - \left(\frac{ik}{z}y_2+T_1\Delta y-\frac{i a_yT_1}{\alpha_1}\right)y'_2\right]{\rm d}x'_2{\rm d}y'_2,
\end{split}
\end{equation}
Using the formula
\begin{equation}
    (a+b)^m = \sum_{n=0}^m C_m^n a^{m-n} b^n,
\end{equation}
where $C_m^n = m!/n!(m-n)!$, we can expand $[ia_x+{\rm sgn}(m) a_y+2T_1(x'_2-i\text{sgn}(m)y'_2)]^{|m|}$ in a binomial series to get
\begin{equation}
\begin{split}
    & \sum_{n=0}^{|m|} C_{|m|}^n [ia_x+{\rm sgn}(m) a_y]^{(|m|-n)}(2T_1)^n\\&\times \iint \left[x'_2-i\text{sgn}(m)y'_2\right]^{n}\left[x'_2+i\text{sgn}(m)y'_2\right]^{|m|} \\
    &  \times\exp\left[-\left(\alpha^*_1-\frac{T_1^2}{\alpha_1}\right)\left(x'^2_2+y'^2_2\right) - i \xi_x x'_2 - i\xi_y y'_2\right]{\rm d}x'_2{\rm d}y'_2,
    \label{step}
\end{split}
\end{equation}
where we have used
\begin{subequations}
\begin{align}
\xi_x&=\frac{kx_2}{z}-iT_1\Delta x-\frac{a_xT_1}{\alpha_1} ,
\\ \xi_y&=\frac{ky_2}{z}-iT_1\Delta y-\frac{a_yT_1}{\alpha_1}.
\end{align}
\end{subequations}
Now, we can write Eq.~(\ref{step}) in cylindrical coordinates as
\begin{equation}
\begin{split}
    & \sum_{n=0}^{|m|} C_{|m|}^n [ia_x+{\rm sgn }(m)a_y]^{(|m|-n)}(2T_1)^n \\
    & \times \int {\rho'_2}^{|m|+n+1}\exp\left[-\left(\alpha^*_1-\frac{T_1^2}{\alpha_1}\right) \rho^{\prime2}_2\right] \text{d}\rho'_2 \\
    & \times \int \exp\left[{\rm sgn}(m)i (|m|-n) \phi'_2 - i \xi_x \rho'_2\cos\phi'_2 - i\xi_y \rho'_2\sin\phi'_2 \right] \text{d}\phi'_2 
\end{split}
\end{equation}
and further employ the formula \cite{integrals}
\begin{align}
    \int_0^{2\pi}e^{\pm int+i(x{\rm sin}t+y{\rm cos}t)}{\rm d}t=\frac{2\pi J_{\pm n}\left(\sqrt{x^2+y^2}\right)}{(-1)^n\sqrt{x^2+y^2}^n}(x\mp iy)^n,  
\end{align}
to solve the azimuthal angle integral, as well as \cite{integrals}
\begin{subequations}
\begin{gather}
    \int_0^{\infty}x^{2n+u+1}e^{-x^2}J_u(2x\sqrt{z}){\rm d}x=\frac{n!}{2}e^{-z}z^{\frac{1}{2}u}L_n^u(z),\\
    L_n^{\alpha} (x)=\sum_{k=0}^n\frac{(-1)^k}{k!}C_{n+{\alpha}}^{n-k}x^k,
\end{gather}
\end{subequations}
for the radial integral. After some calculations, the product of the the propagated kernels is obtained as
\begin{align}
    & H^\ast(\bm{\rho}_1,v,z)H(\bm{\rho}_2,v,z) = \nonumber \\
    & \hspace{4mm} \exp{\left[-\Theta(x,y,T)-\frac{ik}{2z}\left(\boldsymbol\rho_1^2-\boldsymbol\rho_2^2\right)-T_1\left(\boldsymbol\rho_1-\boldsymbol\rho_2\right)^2\right]}\nonumber\\
    & \hspace{4mm} \times w_m^2(z)\sum_{n=0}^{|m|}\left(\tilde{C}_{|m|}^n\right)^2\left[F(x,y,T)\right]^{|m|-n}.
\end{align}
which is Eq.~(\ref{hh}) in the main text, and the employed shorthand notations are
\begin{subequations}
\begin{align}
\Theta(x,y,T) &= (a_x^2+a_y^2)/(4\alpha_1)+(\xi_x^2+\xi_y^2)/(4\alpha_2), \\
w_m(z) & = z^{|m|} w_0/[k^{|m|}w(z)^{|m|+1}], \\
\tilde{C}_{|m|}^n &= \binom{|m|}{n}\sqrt{n!(4T_1)^n}, \\
F(x,y,T) &= [a_y+{\rm sgn}(m)ia_x][\xi_y-{\rm sgn}(m)i\xi_x]\nonumber\\&-(T_1/\alpha_2) (\xi_x^2+\xi_y^2),\\
\alpha_2 &= \alpha_1^*-\frac{T_1^2}{\alpha_1}.
\end{align}
\end{subequations}
Lastly, the turbulence-corrected, $z$-dependent waist of the beam is given by
\begin{equation}\label{D-17}
w^2(z)= \left(\frac{z}{kw_0}\right)^2 + 4T_1\left(\frac{z}{k}\right)^2 + \left(1-\frac{4\pi v z}{k}\right)^2w_0^2,
\end{equation}
whereas the free space waist is simply
\begin{equation}
w_{fr}^2(z)= \left(\frac{z}{kw_0}\right)^2 + \left(1-\frac{4\pi v z}{k}\right)^2w_0^2.
\end{equation}
Employing these equations together with the main text considerations, allows us to analytically model a self-focusing vortex beam of arbitrary OAM content at any propagation distance.

\begin{acknowledgments}
M. L. acknowledges the financial support of the National Natural Science Foundation of China (NSFC) (61805080), Hunan Provincial Natural Science Foundation of China (2019JJ50366) and the China Scholarship Council (CSC). 

M. K., and M. O. acknowledge the financial support of the Academy of Finland Flagship Programme (PREIN-decisions 320165, 320166). 

C. D. acknowledges the financial support of the National Natural Science Foundation of China (NSFC) (12174171) and 2020 Central Plains Talents Program of Henan (ZYYCYU202012144).
\end{acknowledgments}

\section*{Data Availability Statement}

The data that support the findings of this study are available from the corresponding author upon reasonable request.

\end{document}